\documentclass[conference]{IEEEtran}
\IEEEoverridecommandlockouts

\newif\ifDEBUG
\DEBUGfalse

\newif\ifANONYMOUS
\ANONYMOUSfalse

\usepackage{booktabs} 
\usepackage{amsmath}
\usepackage{algorithm}
\usepackage{algpseudocode}
\usepackage[normalem]{ulem} 
\usepackage{xspace}
\usepackage{multirow} 
\usepackage{balance} 
\usepackage{fancyvrb} 
\usepackage{caption}

\usepackage{enumitem}
\setlist[itemize]{leftmargin=*,noitemsep,topsep=0pt}
\setlist[enumerate]{leftmargin=*}

\usepackage{etoolbox}
\makeatletter
\patchcmd{\@makecaption}
	{\scshape}
	{}
	{}
	{}
\makeatletter
\patchcmd{\@makecaption}
	{\\}
	{.\ }
	{}
	{}
\makeatother
\def\tablename{Table}

\newcommand{\code}[1]{\texttt{{\small #1}}}

\newcommand{\ie}{\textit{i.e.,}\xspace}
\newcommand{\eg}{\textit{e.g.,}\xspace}
\newcommand{\etal}{\textit{et al.}\xspace}

\usepackage{mdframed}
\mdfsetup{skipabove=0.5\topskip,skipbelow=0.5\topskip,align=center}

\usepackage{amsthm}
\newtheorem{thm}{Theorem}

\DeclareMathSymbol{\mlq}{\mathord}{operators}{``}
\DeclareMathSymbol{\mrq}{\mathord}{operators}{`'}

\newif\ifSAVESPACE
\SAVESPACEfalse

\ifSAVESPACE

\else

\fi

\usepackage{todonotes}
\usepackage{soul}
\usepackage{fontawesome}

\ifDEBUG
    \newcommand{\TS}[1]{\todo[color=cyan,inline]{Taylor says:#1}}
    \newcommand{\JD}[1]{\todo[color=yellow,inline]{Jamie says:#1}}
    \newcommand{\GKT}[1]{\todo[color=green,inline]{George says:#1}}
    \newcommand{\WJ}[1]{\todo[color=orange,inline]{Wenxin says:#1}}
    
    \newcommand{\PJ}[1]{\todo[color=green,inline]{Purvish says:#1}}
    
    \newcommand{\TODO}[1]{\hl{#1}}
    \newcommand{\PROMPT}[1]{\hl{#1}}
    
\else
    \newcommand{\TS}[1]{}
    \newcommand{\JD}[1]{}
    \newcommand{\GKT}[1]{}
    \newcommand{\WJ}[1]{}
    \newcommand{\PJ}[1]{}
    
    \newcommand{\TODO}[1]{}
    \newcommand{\PROMPT}[1]{}
\fi

\usepackage{hyperref}

\usepackage{cleveref}
\crefformat{section}{\S#2#1#3}
\crefname{figure}{Figure}{Figures}
\crefname{table}{Table}{Tables}
\crefname{theorem}{Theorem}{Theorems}
\crefname{thm}{Theorem}{Theorems}
\crefname{lemma}{Lemma}{Lemmata}
\crefname{equation}{Eqt.}{Eqts.}
\crefformat{Grammar}{Grammar #1}
\crefname{appendix}{Appendix}{Appendices}
\crefname{listing}{Listing}{Listings}

\usepackage{url}

\usepackage{tcolorbox}

\begin{document}

\title{Reusing Deep Learning Models: Challenges and Directions in Software Engineering}


\ifANONYMOUS
\else

\author{
\IEEEauthorblockN{James C. Davis\IEEEauthorrefmark{1}, Purvish Jajal\IEEEauthorrefmark{1}, Wenxin Jiang\IEEEauthorrefmark{1},\\
Taylor R. Schorlemmer\IEEEauthorrefmark{1}, Nicholas Synovic\IEEEauthorrefmark{2}, and George K. Thiruvathukal\IEEEauthorrefmark{2}} 
\IEEEauthorblockA{
\IEEEauthorrefmark{1}Department of Electrical \& Computer Engineering, Purdue University, West Lafayette, IN, USA\\
\{davisjam,pjajal,jiang784,tschorle\}@purdue.edu}
\IEEEauthorblockA{
\IEEEauthorrefmark{2}Department of Computer Science, Loyola University Chicago, Chicago, IL, USA\\
\{nsynovic,gthiruvathukal\}@luc.edu}
\thanks{All authors are listed in alphabetical order as equal contributors to this manuscript. Please direct correspondence to davisjam@purdue.edu and gthiruvathukal@luc.edu.}
}

\fi

\maketitle

\begin{abstract}
Deep neural networks (DNNs) achieve state-of-the-art performance in many areas, including
  computer vision,
  system configuration,
  and question-answering.
However, DNNs are expensive to develop, both in intellectual effort (\eg devising new architectures) and computational costs (\eg training).
Re-using DNNs is a promising direction to amortize costs within a company and across the computing industry.
As with any new technology, however, there are many challenges in re-using DNNs.
These challenges include both missing technical capabilities and missing engineering practices.

This vision paper describes challenges in current approaches to DNN re-use.
We summarize studies of re-use failures across the spectrum of re-use techniques, including conceptual (\eg re-using based on a research paper), adaptation (\eg re-using by building on an existing implementation), and deployment (\eg direct re-use on a new device).
We outline possible advances that would improve each kind of re-use.

\end{abstract}


\section{Introduction} \label{sec:intro}

\begin{figure*}[ht!]
    \centering
    \includegraphics[width=0.75\linewidth]{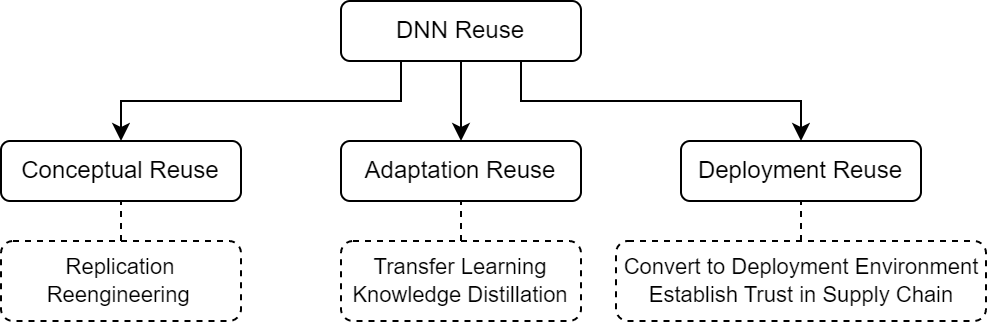}
    \caption{
    Deep neural network reuse is the process of using existing DNN technology for another purpose. 
    We focus on three distinct types: \textit{conceptual reuse}, where existing theory is repurposed; \textit{adaptation reuse}, where existing DNN models are modified; and \textit{deployment reuse} where existing DNN models are converted for use in a new environment.
    Dashed boxes provide examples of each type.
    \JD{@Taylor: There is no link to the figure source in the paper, so I cannot edit it myself. Here is what the boxes should say: ``Replication'' and ``Reengineering'', then (no change to middle box), then ``Convert to deployment environment'' and ``Establish trust in supply chain''}
    }
    \label{figure:reuseParadigms}
\end{figure*} 

The long ``AI Winter''~\cite{mccorduck1991machines} is over.
Machine learning, especially the deep learning paradigm, matches or exceeds human performance in diverse tasks~\cite{SurpassingHumanLevel_he2015delving,hatcher2018survey}.
These tasks include signal processing (\eg
  sight~\cite{goel_survey_2020,goel2022efficient},
  hearing~\cite{nassif2019speech},
  smell~\cite{hines1999electronic}), general reasoning (\eg question-answering~\cite{nakano_webgpt_2022}, synthesizing information), and artistic expression (\eg Dall-E~\cite{ramesh_zero-shot_2021}, music-making~\cite{dhariwal_jukebox_2020}).
Although we expect further scientific breakthroughs in performance, many techniques are mature and already being applied to domains such as marketing~\cite{tkavc2016artificial}, medicine~\cite{anwar2018medical}, and autonomous vehicles~\cite{kuutti2020survey}.
Machine learning has entered the domain of the software engineer. 

One key software engineering technique is \emph{reuse}~\cite{hendrickson2021thinking}.
To reduce engineering costs, software engineers recognize opportunities to reuse, and they also determine an appropriate reuse paradigm.
Software reuse has been exhaustively studied in traditional software engineering,
  both conceptually (\eg~\cite{krueger1992software,frakes2005software}) 
  and 
  empirically (\eg~\cite{sadowski2015developers,abdalkareem2017developers,michael2019RegexesareHard}).
However, for deep learning software the study of reuse has just begun.

Deep learning (DL) is an emerging field whose probabilistic nature and data-driven approach differs from traditional software~\cite{Karpathy2017Software2.0}.
The engineering community lacks long-term experience in appropriate engineering methods for deep learning~\cite{Lecun2015}.
Prior work has characterized the challenges of \emph{developing} DL software~\cite{AmershiMicrosoft2019SEforML,Rahman2019MLSEinPractice}.
Yet relatively little is known about DL \emph{reuse} and the specific challenges engineers face when trying to reuse these models.

In this vision paper, we examine challenges and future directions for the reuse of deep neural networks (DNNs). 
\cref{figure:reuseParadigms} depicts the three paradigms of DNN reuse that we consider: conceptual reuse, adaptation reuse, and deployment reuse.
\textit{Conceptual reuse} is the replication of DNN techniques from sources like academic literature.
\textit{Adaptation reuse} is the modification of pre-trained DNNs to work in a particular use case.
\textit{Deployment reuse} is the conversion of pre-trained DNNs into forms which operate in different environments.

The paper proceeds as follows.
In~\cref{sec:background}, we discuss the general landscape of work on DNNs.
We define the different types of reuse.
In~\cref{sec:challenges}, we go into detail about each type of reuse, focusing on 
  the general nature of the challenge and
  prior work that clarifies it.
In~\cref{sec:directions}, we focus on
  open problems worthy of community attention.
The scope of this work is to introduce the current conceptual, adaptation, and deployment challenges facing DNN reuse, with a discussion of future directions to address these issues.  
We do not aim to address the (many!) other software engineering challenges related to DNNs, such as when to use them, for what purposes, how to integrate them into existing software systems, and how to debug them.

We hope the community's continued efforts to understand the software engineering implications of working with DNNs will lead to standardization of software engineering practices and the development of tools, both of which were greatly helpful to the advancement of traditional software development. 

\section{Background} \label{sec:background}


\subsection{Engineering Deep Neural Networks} \label{sec:background-EngineeringDNN}

\subsubsection{DNN Concepts}

Software that implements a DNN has three main components:
  the DNN model itself (a parameterized computational graph);
  a data pipeline (preprocessing to manipulate data into an appropriate format);
  and
  the training mechanisms~\cite{banna2021experience}.
Due to the computational costs of training and inference, DNN software is also optimized.
We discuss background for each.

\begin{figure*}[t]
    \centering
    {\includegraphics[width=0.93\linewidth]{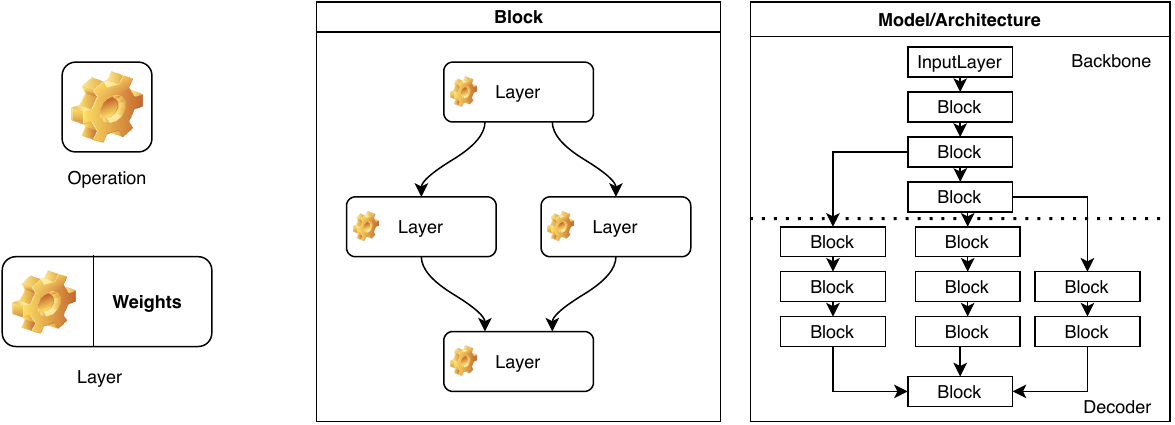}}
    \caption{
    Components of a deep neural network (DNN), represented at different levels of abstraction.
    A DNN is a composition of weighted operations.
    These are combined into a \emph{layer}; a group of layers into a \emph{block}; and a group of blocks into a sub-graph such as a ``backbone'' or a ``head''.
    }
    \label{fig:modelComponentsByHierarchy}
\end{figure*}

\paragraph{The DNN Model}
A DNN model is a parameterized computational graph~\cite{Lecun2015}, \ie the composition of weighted operations~\cite{goodfellow2016deeplearning}. 
Layers can be grouped into blocks based on their purpose or frequency of occurrence.
For example, Long Short-Term Memory (LSTM) units are commonly used blocks in recurrent neural networks, and they are particularly designed to handle time-series data~\cite{hochreiter1997LSTM}.
Each layer and block can be designed and tested individually~\cite{Braiek2020onTestingMLPrograms, Li2022TestingMLSystemsinIndustry}, and then integrated into one structure prior to training.
These model sub-elements are shown in~\cref{fig:modelComponentsByHierarchy}.
The layer sizes must be noted to maintain the shape of the model and ensure that adjustments are made so that the output of one element is the same size as the input of the next.

The architecture of a DNN model will vary based on its purpose.
As an example, in the computer vision domain, typical models have three main components: feature extractor (backbone), decoder~\footnote{There exists some terminological confusion within the literature on ``decoder''.  A \textit{model decoder} refers to the post-processing component of the model. The \textit{data pipeline decoder} transforms data into a format that is directly usable by downstream operations.},
and head~\cite{carion2020end2endODwithTransformers}.
The backbone refers to the combination of layers responsible for feature extraction. 
The backbone is usually a large number of sequential layers and the majority of the model training process is to adjust the backbone parameters. 
The decoder, which usually has less layers than the backbone, handles post-processing and serves as a connection between the backbone and the head. 
The head, often the smallest element, is responsible for the final model task (ex: image classification). 
Depending on the model architecture, the components included within the backbone and decoder may vary.

\begin{figure}
    \centering
    \includegraphics[width=0.87\columnwidth]{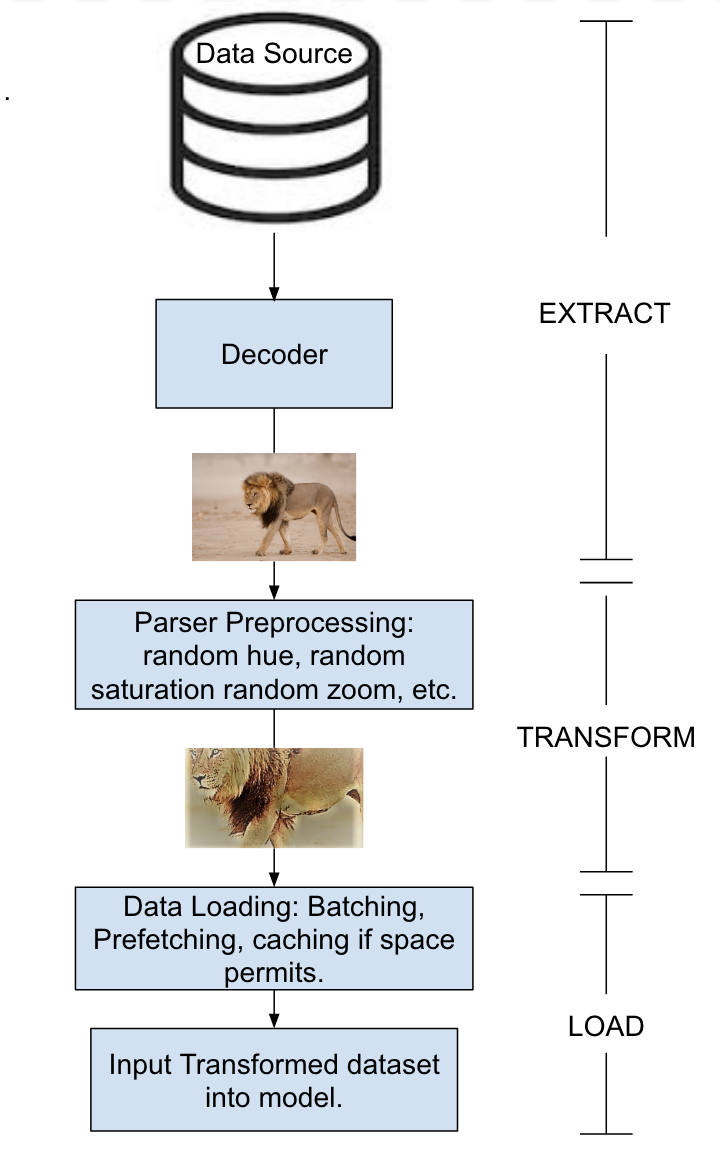}
    \caption{
    Illustration of a data pipeline following the Extract-Transform-Load design pattern.
    The specific pipeline is for the You-Only-Look-Once (YOLO) model family~\cite{Redmon2018YOLO}.
    }
    \label{fig:data-pipeline}
\end{figure}

\paragraph{The Data Pipeline}
A data pipeline
  extracts raw data such as pictures from a storage source (\emph{decoder}),
  transforms the data into the required input format (\emph{parser}),
  and loads the transformed data into the GPU (\emph{loader}).
The Extract-Transform-Load (ETL) architecture~\cite{Wiki:ETL} is a common pattern for data pipelines as it simplifies implementation.
It also promotes reuse: different models can share datasets (Extract), vary in their data augmentations (Transform), and have similar final stages (Load).
An example data pipeline is depicted in~\cref{fig:data-pipeline}.

The data pipeline may also be modified as part of training.
Some changes are for robustness, while others are for performance~\cite{Shorten2019SurveyonImageDataAug}.
For robustness, \emph{data augmentations} can be added to ensure that a model can perform well not only on well formed input but also on more unusual input~\cite{Yokoyama2020BuildingRobustDNNApplicationsAnIndustrialCaseStudyofDataAug}.
For example, in computer vision, images may be rotated, their colors may be changed, and ``filters'' may be added such as fog~\cite{yin2019CVRobustness}.
For performance, the data processing may be optimized to avoid bottlenecks during training.
If a dataset fits in the available system memory then it can be cached.
Else, the dataset must be processed in batches.


\paragraph{The Training Mechanisms}
The final aspect of a DNN is its training --- the identification of parameters for the DNN's operations (``weights'') that yield acceptable functionality~\cite{Lecun2015, goyal2017accurate}.
When training, engineers must provide a loss function.
A loss function defines the learning curve of the DNN and is used to adjust the network weights (\eg via inspecting the loss gradient over different parameter changes).
Once the weights are finalized by minimizing the loss function, the loss function itself is no longer required.

\paragraph{Optimizations}
Based on the device being used for training, the model training time and efficiency will vary drastically~\cite{sze2017EfficientProcessingofDNNs,dean2012LargeScaleDistributedDNN}.
Certain measures can be taken to increase the training efficiency on average.
For example, the data pipeline operations have a strong effect on the training time.
Thus, the operations used during the data pipeline stage must be efficient and based on built-in operations whenever possible (\eg vectorized mathematics rather than \code{for}-loops).
Training hyper-parameters such as the number of epochs may be adjusted depending on the training device, (\eg whether it is a CPU, GPU or TPU).

However, optimizing the data pipeline may violate the modular design of the data pipeline.
Breaking module boundaries is a common effect of performance optimization~\cite{Shen2023NCMAs}.
For example, one common data augmentation applied while training a computer vision model is image blurring.
Image blurring uses a convolution operation which is slow on a CPU but efficient on a GPU.
Transitioning to a GPU only for one operation and then back to the CPU would be time-consuming, but this issue can be tackled by placing the blurring operation at the end of the data pipeline.
This is because the blur operation would immediately be followed up by the DNN training process on the same device.  
Thus, the order of operations can influence the model's training time, and they may be re-arranged in an implementation for efficiency.

\subsubsection{Distributing a DNN Model for re-use}

DNN software is sometimes shared, either open-source or as binaries.
The model may be shared alone, or with weights after training.
A DNN model with both architecture and pre-trained weights is called a \emph{Pre-Trained Model (PTM)}~\cite{Han2021PTM}.
When a PTM is shared for re-use in an organized manner, it is called a \emph{pre-trained model package}~\cite{Jiang2022PTMReuse}.
The package may include the PTM as well as the names of the authors, the bill of materials (SBOM), license information, versioning, test suite, and so on.

\subsubsection{Engineering ``Best Practices''}

Major technology companies have shared different kinds of studies on machine learning (ML) best practice (\eg Google~\cite{Breck2017aRubricforMLProductionReadinessandTechnicalDebtReduction}, Microsoft~\cite{AmershiMicrosoft2019SEforML}, and SAP~\cite{Rahman2019MLSEinPractice}). Google~\cite{Breck2017aRubricforMLProductionReadinessandTechnicalDebtReduction} and Microsoft~\cite{AmershiMicrosoft2019SEforML} provide high level guidelines on the current state of the ML models and potential improvements. 
Breck \etal from Google present 28 quantified tests as well as monitoring needs and propose an easy-to-follow rubric to improve ML production readiness and reduce technical debt~\cite{Breck2017aRubricforMLProductionReadinessandTechnicalDebtReduction}. 
Rahman \etal present a case study on SAP~\cite{Rahman2019MLSEinPractice}.
They discuss the challenges in software engineering, ML, and industry-academia collaboration and specifically point out the demand for a consolidated set of guidelines and recommendations for ML applications.
Unlike the guidance on high-level architectures and organizational processes shared by Google and Microsoft, we focus on lower-level engineering and programming patterns. 

In addition to views of the industry, there are also works on software engineering practice for the engineering of DL software from academic perspectives~\cite{Zhang2019empiricalstudyofcommonchallengesindevelopingDLApplications, Zhang2019SEPracticeintheDevelopmentofDLapplications,Serban2020AdoptionandeffectsofSEBPinML,Washizaki2019a}. 
Zhang \etal present 13 valuable findings in the practice of the engineering life cycle of DL applications~\cite{Zhang2019SEPracticeintheDevelopmentofDLapplications}. 
They indicate that we are in great demand for test cases which could be used to check the correctness of the system functions.
Based on their findings, they give suggestions to both practitioners and researchers including using well-known frameworks, improving the robustness of DL applications, and proposing new bug locating tools and logs.
Serban \etal conducted a survey and quantified practice adoption and demographic characteristics, then propose a positive correlation between practices and effects~\cite{Serban2020AdoptionandeffectsofSEBPinML}.
However, there has been limited focus on perspective of reusability in the context of DL models. In this vision paper, we present a comprehensive summary of existing reuse paradigms, and propose challenges as well as future directions in the field of software engineering

\subsection{Reuse Paradigms} \label{sec:background-reuse}

We identify three major reuse paradigms for deep learning models.
These reuse paradigms are related to those proposed by Sommerville~\cite{Sommerville2016SEText} and Krueger~\cite{krueger1992software} for reuse in traditional software engineering, but take somewhat different forms in deep learning engineering.
Specifically, the reuse of deep learning models necessitates unique considerations including the demand for significant computational resources, the need to accommodate specific hardware configurations, and the inherent dependencies on diverse datasets.

\begin{itemize}
    \item \textbf{Conceptual Reuse:} \emph{Replicate and reengineer the algorithms, model architectures, or other concepts described in academic literature and similar sources, integrating the replication into new projects}. 
    An engineer might do this because of licensing issues or if they are required to use a particular deep learning framework (\eg TensorFlow) but are reusing ideas previously realized in another deep learning framework (\eg PyTorch)~\cite{Intro2TFMG}.
    This paradigm is related to Sommerville's notion of ``abstraction reuse'', where an engineer reuses knowledge but not code directly.
    This paradigm is also related to reproducibility in the scientific sense, since an engineer independently confirms the reported results of a proposed technique~\cite{Huston2018AIfacesReproducibilityCrisis, Alahmari2020RepeatabilityofDLModels}.
    
    \item \textbf{Adaptation Reuse:} \emph{Leverage existing DNN models and adapt them to solve different learning tasks}.
    An engineer might do this using several techniques, such as transfer learning~\cite{transferlearning} or knowledge distillation~\cite{Hinton2015KnowledgeDistilling}.
    This form of reuse is suitable if there is a publicly available implementation of an appropriate model (a pre-trained model).
    This paradigm is related to Sommerville's notion of ``object/component reuse'', since an engineer must identify existing models suited for a purpose and then customize them for a different task.
    
    \item \textbf{Deployment Reuse:} \emph{Convert and deploy pre-trained DNN models in different computational environments and frameworks}.
    This form of reuse is suitable if there is a perfect fit for the engineer's needs, viz. a DNN trained on the engineer's desired task (\eg demonstrating proof of concept in a hackathon).
    This paradigm is related to Sommerville's notion of ``system reuse'', since an engineer is reusing an entire model (including its training) and deploying it in the appropriate context.
    Deployment often requires the conversion of a DNN from one representation to another, followed by compilation to optimize it for hardware.
\end{itemize}

\vspace{0.1cm}
\noindent
These reuse paradigms are orthogonal.
Multiple forms of reuse are possible in a single engineering project.
For example, an engineering team might
  develop their own version of a decision-making component (conceptual reuse);
  re-use the implementation of a feature extractor DNN as a backbone (adaptation reuse);
  and after training, convert their model to a specialized hardware environment (deployment reuse).

\section{Challenges} \label{sec:challenges}

In this section, we describe the open software engineering challenges associated with deep learning model reuse.
Each subsection addresses one of the three types of reuse: conceptual, adaptation, and deployment. 

\subsection{Challenges in Conceptual Reuse of DNNs} \label{sec:challenges-conceptual}


Conceptual reuse in deep learning primarily takes two forms. 
The first is reproducing results using the same code and dataset. 
The second form is replication and model reengineering, where the same algorithm is implemented (possibly with changes) in a new context, \eg a different deep learning framework or a variation on the original task. 
Each of these forms presents its own set of potential challenges. 

\subsubsection{Reproducibility of Results}
Conceptual reuse of DNN necessitates an understanding of state-of-the-art DNN architectures and algorithms. As part of this process, both engineers~\cite{DeterminedAIReproducibility, FacebookAIReproducibility} and researchers~\cite{MLReproducibilityChallengeSpring2021, CMUMLReproducibility} regard the reproduction of reported DNN results as essential for enhancing comprehension and trust.

Reproducibility is considered a key quality of machine learning software, with a particular emphasis on the thorough exploration of experimental variables and the requirement for comprehensive documentation to achieve reproducibility~\cite{Pineau2020}. Despite this recognition, achieving DNN reproducibility remains a challenging task and continues to be a focal point within the research community~\cite{Huston2018AIfacesReproducibilityCrisis, Pineau2020, Gundersen2018ReproducibilityinAI, Gundersen2018onReproducibleAI, Chen2022TowardsTrainingReproducibleDLModels}.

Many state-of-the-art models are publicly available, but they often exist in the research prototype stage. This stage is typically characterized by an absence of rigorous testing, inadequate documentation, and a lack of considerations for portability. These factors contribute to the ongoing ``reproducibility crisis'' in the field of deep learning~\cite{Breck2017aRubricforMLProductionReadinessandTechnicalDebtReduction, FacebookAIReproducibility}.

\subsubsection{Model Replication and Reengineering}

Replicating and reengineering DNNs is tricky, even when referring to the original code of the research prototypes~\cite{banna2021experience}.
Deep learning frameworks contain many sources of variability that can limit replicability~\cite{Pham2020AnalysisofVarianceinDLSWSystems,pham_deviate_2021}.

Jiang \etal described three challenges of DNN replication and reengineering:
  model operationalization, portability of DL operations, and performance debugging~\cite{Jiang2023CVReengineering}.
First, the \textit{model operationalization}, or the difficulty in choosing and validating the correct model for the task, can be confusing for reusers. 
It can be hard to distinguish between multiple implementations, to identify the implemented conceptual algorithm(s), and to evaluate their trustworthiness.
Second, the \textit{portability of deep learning operations} is inconsistent across frameworks. 
Some deep learning operations only support certain hardware which makes it harder to transfer the implementation to frameworks and environments.
Consequently, \textit{performance debugging} becomes necessary. This step, however, poses its own set of challenges for engineers because of the stochastic nature of deep learning systems~\cite{Rahman2019MLSEinPractice}.

\begin{figure}[t]
    \centering
    \includegraphics[width=0.60\columnwidth]{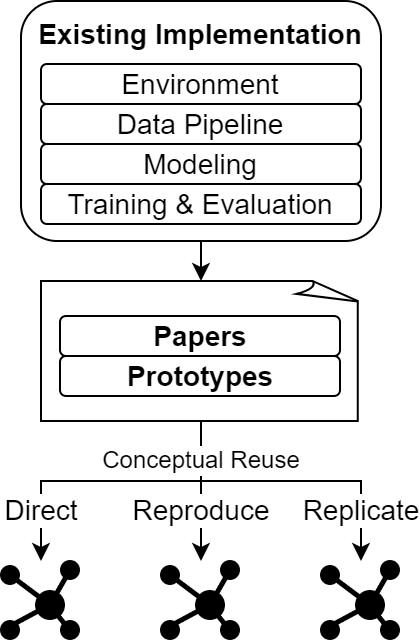}
    \caption{
        Overview of the conceptual reuse of deep learning models.
        An engineer learns about deep learning ideas from existing implementations, \eg research papers and prototypes.
        They use this to guide their own development of a deep learning model,
          \eg implementing it directly,
          reproducing the model,
          or
          replicating the model.
    }
    \label{figure:ConceptualReuse}
\end{figure}


\subsection{Challenges in Adaptation Reuse of DNNs} \label{sec:challenges-adaptation}



Software engineers use multiple techniques to modify existing DNNs for use in their specific applications.
Even though adapting a model is more efficient than training a model from scratch, engineers still face challenges with adaptation.
Software engineers face both \textit{technical} (with respect to an adaptation technique) 
and \textit{decision-making} (with respect to the engineering process)
challenges when adapting a DNN to a particular use case.
This section explores these challenges.

\paragraph{Technical Adaptation Challenges}

\textbf{Techniques}
Existing deep learning models can be adapted by engineers to new tasks through various techniques, such as transfer learning, dataset labeling, and knowledge distillation~\cite{transferlearning, Han2021PTM, datasetlabeling, knowledgedist}. 
The practice of reusing DL models as pre-trained models (PTMs) reduces the need for users to train their own models, thereby facilitating rapid model development for an expanded range of tasks~\cite{Han2021PTM, marcelino_transfer_2022}. 

\begin{figure}[t]
    \includegraphics[width=0.95\columnwidth]{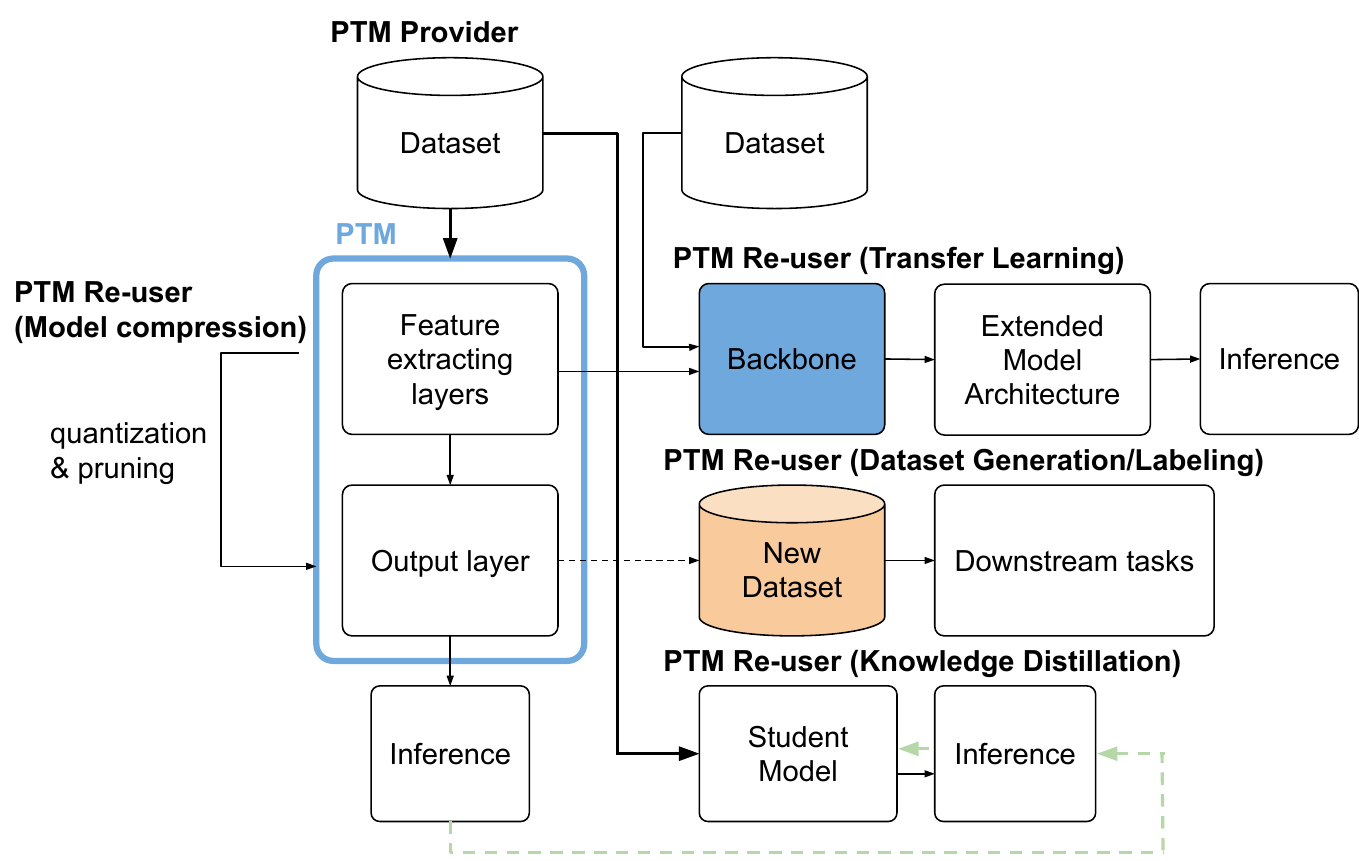}
    \caption{
    A pre-trained deep neural network model (PTM) can be adapted for different tasks via transfer learning, quantization and pruning, and knowledge distillation. 
    This figure is reused from~\cite{Jiang2022PTMSupplyChain} with their permission.
    }
    \label{fig:AdaptationReuse}
\end{figure}

\cref{fig:AdaptationReuse} shows multiple methods for PTM reuse. 
A PTM provider will train a deep learning model on a dataset, resulting in a DNN checkpoint that includes specific architecture and weight parameters. 
A PTM user can then employ one of the aforementioned methods to adapt this model to a new context or application. 
The resulting model may be more compact, easier to manage, and more resource-efficient~\cite{Han2021PTM, knowledgedist}.


\textbf{Accuracy and Latency:}
Engineers adapt DNNs to different hardware constraints and environments.
For example, the adaptation of DNNs to embedded devices (also known as IoT or Edge devices) often demands significant engineering efforts, encompassing techniques such as model compression and hardware acceleration~\cite{Deng2020ModelCompressionandHWAccelerationforNN}.
Engineers report that they are unaware of ``push-button'' techniques to adapt DNNs across hardware environments~\cite{Gopalakrishna2022IoTPractices}.

\textbf{Fairness and Robustness}
In addition to accuracy and latency, engineers reusing Deep Neural Networks (DNNs) may also consider properties such as fairness and robustness~\cite{pessach2022FairnessinML, mehrabi2021BiasandFairnessinMLSurvey, carlini2017RobustnessNN}. A multitude of strategies has been developed to boost the fairness of DNNs, encompassing local interpretability techniques~\cite{sundararajan2017axiomatic, ribeiro2016MLInterpretability} and model-agnostic methods~\cite{Gesi2023LeveragingFeatureBias}.
Regarding robustness, recent research has shown that decomposition techniques can contribute to the resilience of DL systems~\cite{pan2020decomposing, pan2022decomposing}. By enabling model modularization, these techniques provide engineers with the flexibility to substitute faulty or inefficient components during the adaptation reuse process~\cite{imtiaz2023decomposing}. This approach underscores the potential benefits of modular design in bolstering the robustness and maintainability of DL systems in diverse deployment scenarios.
Nevertheless, enhancing fairness and reducing algorithmic bias remain considerable challenges that require further exploration and advancement~\cite{pessach2022FairnessinML}.

\paragraph{Decision-Making Challenges}
Adapting DNNs (\eg as PTMs) is also challenging because of the complicated decision-making process and its costly evaluation loops.
Software engineers must assess if they can reuse a DNN for their task, select a model, adapt the model, evaluate the model, and then deploy the model~\cite{Jiang2022PTMReuse}.
Several challenges exist for this decision-making process --- they include: model selection, discrepancies, and security and privacy risks~\cite{Jiang2022PTMReuse}.

First, selecting an appropriate DNN for reuse is a difficult process for software engineers.
While model registries, such as Hugging Face, are popular in the deep learning community, they often lack infrastructure that provides attributes helpful to the reuse process.
These attributes would include DNN information like model provenance, reproducibility, and portability on top of traditional software information like popularity, quality, and maintenance~\cite{Jiang2022PTMReuse}.
On the other hand, missing attributes lead to an increased engineering effort and an expensive evaluation process during model selection.

Next, potential performance discrepancies may arise from poor documentation or non-robust models.
For example, Montes \etal~\cite{Montes2022DiscrepanciesAmongPTNN} demonstrate that popular models can have significant accuracy and latency discrepancies in their implementations in different model hubs.
Jiang \etal~\cite{Jiang2022PTMReuse} also note the lack of documentation for DNNs on Hugging Face.
These issues make it hard for engineers to adapt models to their tasks.
Engineers must spend additional effort discovering different or undocumented behaviors for the DNNs they choose to reuse.

Lastly, attacks against DNNs pose security and privacy risks to products that rely on them.
Because of the unique aspects of machine learning, new types of attacks have emerged specifically for DNNs.
These attacks can target several aspects of the DNN engineering processing including datasets, model parameters, and even model behaviors.
Many of these attacks take advantage of poor documentation, inadequate security features, and performance discrepancies common among open-source DNNs~\cite{Montes2022DiscrepanciesAmongPTNN}.
Traditional software supply chain threats also apply.
Software supply chains contain actors, artifacts, and operations which interact to create some final software product~\cite{okafor_sok}.
When DNNs are reused, they create their own supply chain consisting of models, datasets, maintainers, and training processes~\cite{Jiang2022PTMSupplyChain} --- see~\cref{tab:comparison}.
Actors play a similar role in both supply chain types, but DNN artifacts and operations differ because of the nature of machine learning.

\begin{table}[]
  \centering
  \begin{tabular}{@{}cll@{}}
    \toprule
    \textbf{Aspect} & \textbf{Software Supply Chains} & \textbf{DNN Supply Chains} \\
    \midrule
    \multirow{3}{*}{Actors}         & Software Engineers            & Software Engineers \\
                                    & Testers                       & Testers \\
                                    & Maintainers                   & Maintainers \\
    \midrule
    \multirow{3}{*}{Artifacts}      & Source Code                   & Source Code, Datasets \\
                                    & Software Dependencies         & Frameworks, Pre-trained Models \\
                                    & Infrastructure                & Infrastructure \\
    \midrule
    \multirow{3}{*}{Operations}     & Development                   & Data Collection, Development \\
                                    & Building, Testing             & Model Training, Testing \\
                                    & Deployment                    & Model Deployment \\
    \bottomrule
  \end{tabular}
  \caption{Comparison of Software Supply Chains and DNN Supply Chains~\cite{Jiang2022PTMSupplyChain} by actors, artifacts, and operations described by Okafor \etal~\cite{okafor_sok}.}
  \label{tab:comparison}
\end{table}

We divide attacks against DNNs into four types:
\begin{enumerate}
    \item \textbf{Train-Time Attacks:} These attacks manipulate the model's training procedure (often through a malicious dataset) to affect its behavior~\cite{pitroppakis_taxonomy_survey_ml_attacks}. 
    A prime example is BadNets~\cite{Gu2019BadNets} where a model is trained on a malicious dataset to illicit improper classifications when later used.

    \item \textbf{Idle-Time Attacks:} These attacks maliciously alter existing DNNs to introduce new behavior. 
    Examples of this type include EvilModel~\cite{Wang2021EvilModel,Wang2022EvilModel2} and StegoNet~\cite{liu_stegonet_2020}.
    These examples implant malware directly into the parameters of models, hiding a malicious payload in a trained DNN without significant drops in performance.
    Another example includes LoneNeuron~\cite{Liu2022LoneNeuron} which injects a malicious function into an existing DNN. 

    \item \textbf{Inference-Time Attacks:} These attacks exploit unexpected input-output pairs or recover confidential dataset or parameter information~\cite{pitroppakis_taxonomy_survey_ml_attacks, miao_ml_based_cyber_attacks}.
    For example, Papernot \etal~\cite{papernot_limitations_dl_adversarial_settings} describe a method to generate adversarial inputs for DNNs;
    Shokri \etal~\cite{shokri_membership_inference_attacks} describe a method for membership inference attacks to reveal sensitive training data;
    and
    MAZE~\cite{kariyappa_maze_2021} is a model stealing attack which recreates the target model without access to its dataset.

    \item \textbf{Traditional Supply Chain Attacks:} These attacks compromise upstream software components with the intent of exploiting downstream DNN vulnerabilities~\cite{okafor_sok, Jiang2022PTMSupplyChain}. 
    Ladisa \etal~\cite{Ladisa2022TaxonomyofAttackonOSSSupplyChains} and Ohm \etal~\cite{Ohm2020ReviewofOpenSourceSSCAttacks} enumerate several classifications of attacks on traditional open-source supply chains.
    These attacks are also applicable to DNN supply chains.
    For example, typo-squatting (whereby an attacker uses similar package names) could confuse DNN users into using the wrong PTM.
\end{enumerate}

\subsection{Challenges in Deployment Reuse of DNNs} \label{sec:challenges-deployment}

Once a deep learning model has been developed, possibly through conceptual or adaptation reuse, it must then be deployed.
This deployment is not simple, so we classify this as a form of reuse rather than just a final step.
Deployment faces its own series of challenges because the development environment of DNNs (for training and testing) may differ substantially from deployment environments.
For this reason, engineers typically face two main types of deployment reuse challenges: interoperability between frameworks and hardware, and trust establishment in DL supply chains.

\paragraph{Deep Learning Interoperability}
The advent of DNNs has brought computing platforms that specialize in accelerating their execution~\cite{DL_comp_surver_Li}, consequently, there has been a greater demand for tools enabling the deployment of DNNs onto diverse hardware. 
\emph{Deep learning compilers}~\cite{DL_comp_surver_Li} such as TVM~\cite{TVM_Chen}, OpenVINO~\cite{OpenVINO}, and Glow~\cite{Glow_Rotem_Fix} aim to bridge this gap by taking DL models and converting them to optimized binary code targeting the multitude of hardware available.
Prior work has focused on developing~\cite{TVM_Chen, Glow_Rotem_Fix} and testing DL compilers~\cite{NNSmith_Liu, Metamorphic_DL_Compilers_Xiao}, as well as empirical studies of DL compiler failures~\cite{comp_DL_comp_bugs_Shen}. 

\begin{figure}[b]
    \includegraphics[width=0.98\columnwidth]{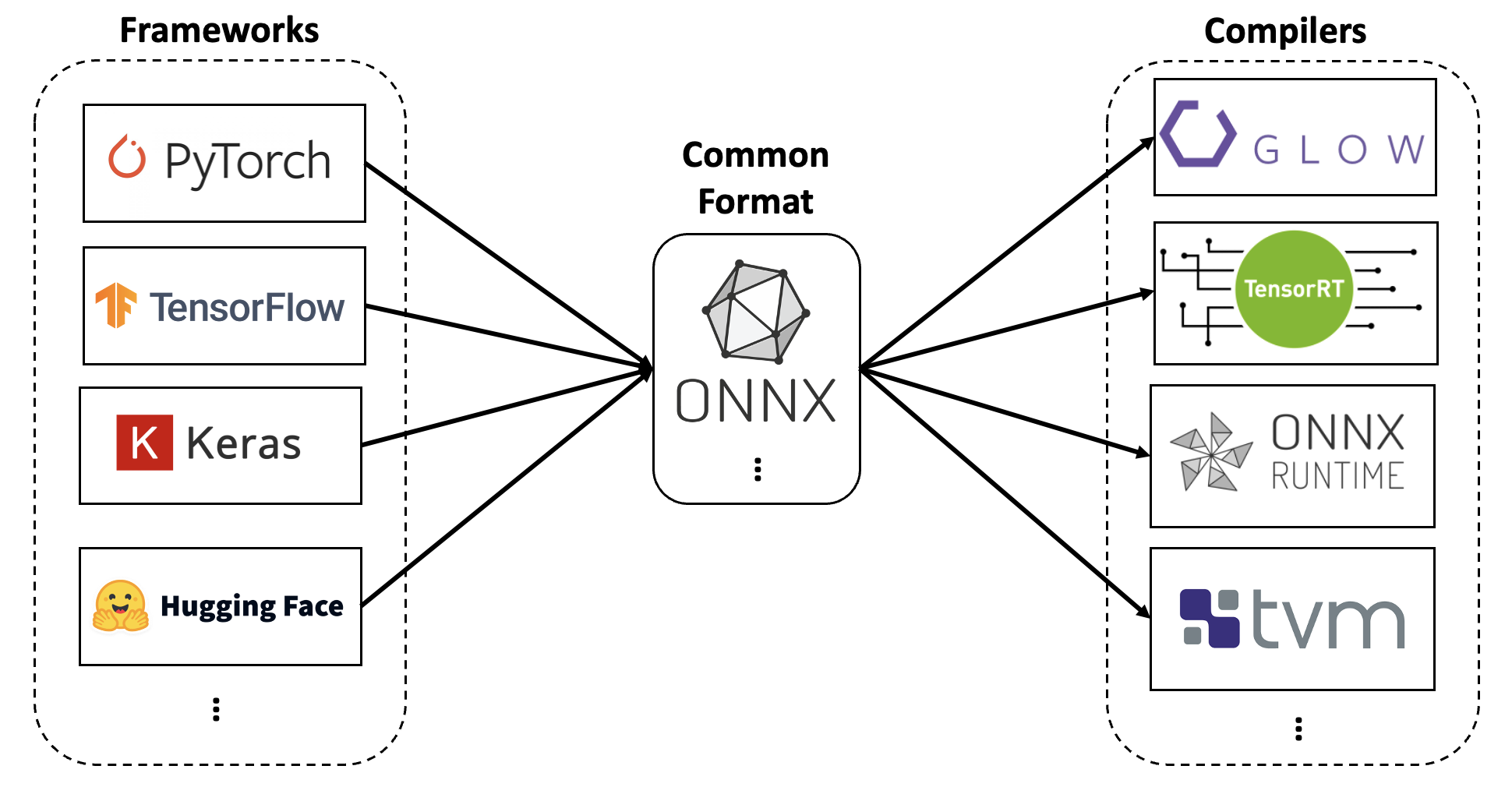}
    \caption{A common format like ONNX (Open Neural Network eXchange) is used as an intermediary to adapt DNN written with general-purpose DL frameworks so that it works on hardware-specific DL compilers. 
    }
    \label{fig:interop}
\end{figure}

The introduction of DL compilers largely addresses the challenges of deploying to diverse hardware but it introduces challenges in the use of compilers. 
Many compilers do not support all frameworks thus the interoperability of frameworks and compilers poses a challenge in the deployment of DNNs.
To address this common representations such as ONNX act as intermediaries between DL frameworks and compilers~\cite{ONNXFixed}.
\cref{fig:interop} depicts how common representations can act as intermediaries between frameworks and compilers.
Model converters are used to convert between framework representations and ONNX such as \textit{tf2onnx}~\cite{tf2onnx}.

Prior work on interoperability has largely focused on common representations or model conversion.
Common representations such as ONNX~\cite{ONNXFixed} and NNEF~\cite{NNEF} have been introduced to further interoperability.
Model converters like MMDnn~\cite{enchanceInterop} allow for faithful conversions between frameworks, or allow conversion to and from common representations. 
Though this type of software is largely understudied in a DL context.
Studies on converters have largely focused on 
DNN properties after conversion~\cite{ConvertingDLempirical}.
ONNX conformance test generation has been proposed to ensure ONNX implementations match the ONNX specification~\cite{Sionnx_Cai_Zhou_Ding_Chen_Zhang_2019}.



Recently, model converter failures have been studied in the context of ONNX~\cite{Jajal_Jiang_Tewari_Woo_Lu_Thiruvathukal_Davis_2023}.
It was found that model converters exhibit two common failure symptoms: crashes and wrong outputs. 
Crashes are largely due to the converter being unable to convert operators of the DNN to ONNX. 
This can be due to the converter not yet implementing this conversion, or ONNX not supporting the operator. 
Wrong outputs are when a successfully converted model is not semantically equivalent to the original model.
Such failures suggest that engineers should weigh the risks of model conversion against potential benefits.

\paragraph{Establishing Trust in DL Supply Chains}
Establishing trust in traditional software products is a difficult task~\cite{cybersecurity_enisa_2021}.
This is no different for DNNs.
In \cref{sec:challenges-adaptation} we discuss the wide array of challenges engineers face when attempting to adapt existing DNNs to solve different tasks.
Specifically, we mention many of the attacks which threaten how engineers can reuse models and that the reuse of DNNs creates a supply chain structure.
These attacks make the decision-making process difficult when adapting DNNs, but they also make it difficult for engineers to establish trust in the DNNs they are about to deploy.
In other words, once an organization decides to release a DNN into the wild, they have a hard time making sure users can trust it.


\begin{figure*}[t]
    \centering
    \includegraphics[width=0.8\textwidth]{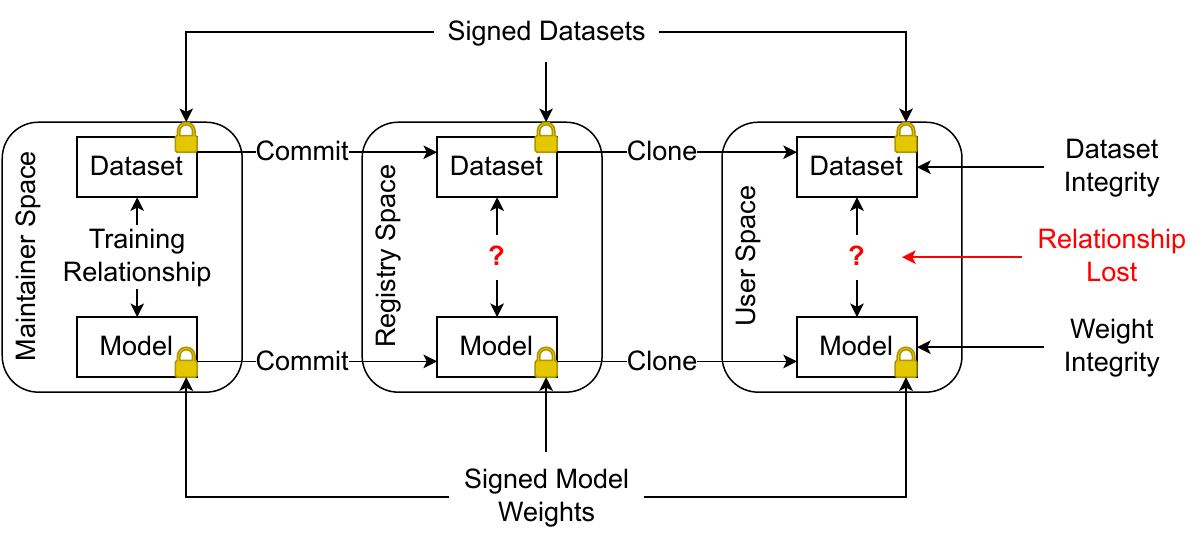}
    \caption{
    Maintainers can sign datasets or models before commiting those files to a registry.
    Attaching signatures to these files enables validation of files between trust domains.
    This enables users to verify dataset and model integrity.
    These signatures only validate the source of a model or dataset --- any relationship between the two cannot be validated.
    Users must trust that maintainers have trained a DNN as described in documentation.
    There is no known way to validate this with high confidence.
    }
    \label{fig:inadequate_ml_signing}
\end{figure*}

Novel characteristics of the DNN supply chain introduce new methods for attackers to degrade trust~\cite{Jiang2022PTMSupplyChain, gu_badnets_2019}.
Furthermore, users are often either unwilling or unable to check for these attacks~\cite{Liu2022LoneNeuron}.
This means users must blindly trust the DNNs they download from open-source registries.
Although traditional security features such as software signing may help to verify file integrity, many attacks slip by these simple measures.
As an example, \cref{fig:inadequate_ml_signing} illustrates a common dependency structure for DNNs that makes it hard for users to validate DNN models and datasets at deployment time (see caption).
Cryptographic methods~\cite{newman_sigstore, intoto} exist to verify that particular files have not changed through deployment, but the relationships between files --- and in particular, DNN dependency relationships --- are not easily verified in this way.
Traditional software dependencies can be verified by checking dependencies through methods like Software Bills of Materials (SBOMs) and reproducible builds~\cite{cisa-sbom, lamb_reproducible_2022}.
Similar techniques are more difficult for DNNs because of non-determinism and training costs~\cite{chen_towards_2022}.

\section{Directions} \label{sec:directions}

\subsection{Directions in Conceptual Reuse of DNNs} \label{sec:directions-conceptual}

Comparing the conceptual reuse of DNNs to that of traditional software~\cite{singh2019OSSreengineering, bhavsar2020MLreengineering}, DNN reuse is mainly based on research products while traditional software reuse is focused on the outputs from engineering teams.
Consequently, the goal and focus of DNN reuse are different~\cite{Jiang2023CVReengineering}.
Considering the differences of conceptual DNN reuse, we propose several research directions, including promoting reusable artifacts and developing engineering tools.

\subsubsection{Evaluating Artifacts for Reuse}

Although conceptual reuse is primarily focused on methods presented in research papers and technical reports, there is a growing effort to include artifact evaluation to support the claims of conference and journal papers.
Typical artifact evaluation when it comes to machien learning includes a minimum-viable prototype, training and evaluation datasets, and results.
While the focus of artifact evaluation in machine learning is primarily on \emph{reproducibility} of a paper's claims, evaluating the \emph{software engineering} of these accepted artifacts would seemingly be a key ingredient of ensuring that a research artifact has some hope of being reproduced by others at the conceptual level (and beyond).
As more and more publication venues incentivize artifacts, an empirical software engineering study to evaluate the software engineering process across these venues, specifically in support of reuse, is a topic worthy of further study.
Separately, we would also encourage artifact evaluation to include a checklist of minimal expectations, which is already a practice in software-focused journals such as the Journal of Open Source Software.



\subsubsection{Testing Tools} \label{Tools}

Testing tools and frameworks are among the high-impact practices in software engineering and greatly improve the reuse potential of software libraries in general.
Testing can also help to improve the DNN reuse process.
Researchers have developed some tools for automated deep learning  testing~\cite{Pei2017DeepXplore}.
Despite these advancements, there is a noticeable lack of adoption of specialized testing techniques for conceptual DNN reuse.

We urge the research community to consider the application of comprehensive testing tools for conceptual DNN reuse tasks.
Since conceptual reuse includes comprehending documentation such as research papers, checklists and tools for extracting knowledge would be helpful.
Validation tools might leverage pre-existing model implementations.
For instance, emerging fuzzing technologies could use adversarial inputs to verify the accuracy of early-stage training, using the original model as a benchmark.
Refining unit testing methods could also reduce overall costs. We encourage exploration of strategies for effective unit and differential testing in DNN software.


\subsection{Directions in Adaptation Reuse of DNNs} \label{sec:directions-adaptation}
Adapting DNNs presents challenges, including complicated decision-making processes, costly evaluation loops, and potential security and privacy risks.
We describe several research directions that support adaptation reuse:
  large-scale model audit,
  infrastructure optimization,
  recommendation systems,
  and
  attack detection. 

\subsubsection{Model Audit}
Prior work found that the trustworthiness of DNNs are concerning due to the lack of DNN transparency. Future work can measure attributes of DNNs that are not currently accessible to engineers. For example, DL-specific attributes can be extracted from provided documentation, source code, and metadata~\cite{Jiang2022PTMReuse}.
Potential risks could be identified by measuring the discrepancies among different DNN models.
These measurements could largely improve the transparency of DNNs and therefore facilitate better adaptation reuse.

\subsubsection{Infrastructure Optimization}
Another challenge of adaptation reuse is the model selection.
Engineers struggle to compare different DNNs and identify a good way to adapt to their downstream task. To facilitate the adaptation, researchers can identify different approaches to support the model selection process.
For instance, providing enhanced documentation, similar to the badges used by GitHub, could offer greater transparency about the capabilities and limitations of a model. This could facilitate a more informed model selection process.

Standardization tools can also play a pivotal role. For example, the use of large language models (LLMs) could facilitate the extraction of metadata for model cards, enabling a systematic comparison of models' performance and requirements. Additionally, a standardized model interface or API could be developed. This would provide a uniform way to interact with and evaluate different models, thereby simplifying the model selection process.

\subsubsection{PTM Recommendation Systems}
Existing literature highlights the challenges associated with technical adaptation and the resulting decision-making process to select a suitable starting point (\cref{sec:challenges-adaptation}).
Numerous studies have sought solutions to the problem of PTM selection.
For instance, You et al. proposed a transferability metric called \textit{LogME} that is used to assess and rank a list of models~\cite{You2021LogME, You2021RankingandTunignPTMs}. Certain works have also employed deep learning techniques to enhance the AutoML process~\cite{Ozturk2022ZeroShotAutoMLwithPTMs, Lu2019AutoDNNSelection4EdgeInference}. Despite these advancements, open-source PTMs remain under-utilized~\cite{Jiang2022PTMReuse}, suggesting the need for a robust PTM recommendation system to aid engineers in adaptation reuse. We encourage researchers to consider various factors during the adaptation reuse process, including diverse fine-tuning approaches, necessary engineering efforts, and the trustworthiness of the model.

\subsubsection{Attack Detection Tool}
Our work indicates that specific attack detection tools are currently missing in DL model registries. For example, Hugging Face only uses ClamAV which can only detect traditional malware but not applicable for DL-specific attacks~\cite{Jiang2022PTMReuse}. The potential risks existing in the model registries make the adaptation reuse harder for engineers. Developing detection tools for DL-specific attacks can therefore improve the trustworthiness of DNNs and help engineers easier adapt existing DNNs.
Detection tools for copyright infringement such as DeepJudge~\cite{chen_copy_2022} might be adapted for this purpose.

\subsection{Directions in Deployment Reuse of DNNs} \label{sec:directions-deployment}

In \cref{sec:challenges-deployment} we identified interoperability and trust as key challenges for DNN deployment.
Here, we identify future research directions that may help engineers deploy DNNs more effectively.

Interoperability plays a key role in the deployment of DNNs, namely the use of model converters (ONNX) to connect frameworks to compilers (as shown in~\cref{fig:interop}).
ONNX model converter failures have been studied~\cite{Jajal_Jiang_Tewari_Woo_Lu_Thiruvathukal_Davis_2023}.
Converters often are not compatible with all ONNX operators, moreover, converted models often are not semantically equivalent to their original models.

Based on prior work, we propose the following directions related to enhancing interoperability, converter testing, and increasing DNN supply chain security.
\begin{enumerate}
    \item \textit{Model Converter Testing:} Model converters often produce models that are semantically inequivalent to the original models. 
    Consequently, testing to identify the source of semantic inequivalence in converters in important.
    
    \item  \textit{Supporting Model Converter Engineering:} ONNX model converters suffer from incompatibilities with the evolution of intermediate representation.
    This results in converters needing updates as ONNX updates to ensure the specification is faithfully followed.
    It follows that these efforts must be supported. 
    Specifically, engineering efforts can be better focused by understanding operator popularity, and domain-specific languages can be used to automatically generate converter code.
    Additionally, these efforts can be avoided with the development of a small stable operator set that can represent all possible operators that can be used in a deep learning model~\cite{Liu2020DNNInterop} --- similar to RISC~\cite{patterson2016computer} and JVM~\cite{lindholm2014java}.

    \item \textit{Supply Chain Security for DNNs:} The software engineering community has been working on systems such as TUF~\cite{Samuel_Survivable_Key_Compromise} and Sigstore~\cite{newman_sigstore} to increase the usability and effectiveness of signatures for package managers.
    The community has also began to develop standards such as Supply-chain Levels for Software Artifacts (SLSA)~\cite{noauthor_supply-chain_nodate} and Software Supply Chain Best Practices from the Cloud Native Computing Foundation (CNCF)~\cite{security_technical_advisory_group_software_2021} to help engineers implement appropriate supply chain security measures.
    Similar efforts should be taken to create systems and standards for the DNN community, \eg determining which concepts still apply to DNN environments and which concepts need to be changed.
    For example,~\cref{fig:inadequate_ml_signing} shows how traditional signing methods may be inadequate in DNN supply chains.
    Future work could consider how to preserve the relationship between a particular DNN and the dataset(s) it was trained on.
    This might be accomplished by watermarking~\cite{li2021survey} or some form of limited reproduction, whereby users validate model-dataset relationships via retraining.
    
\end{enumerate}

\subsection{Assessing the Software Engineering Process of DNNs} \label{sec:directions-miningSoftwareRepos}

\begin{figure*}
    \centering
    \includegraphics[width=0.8\textwidth]{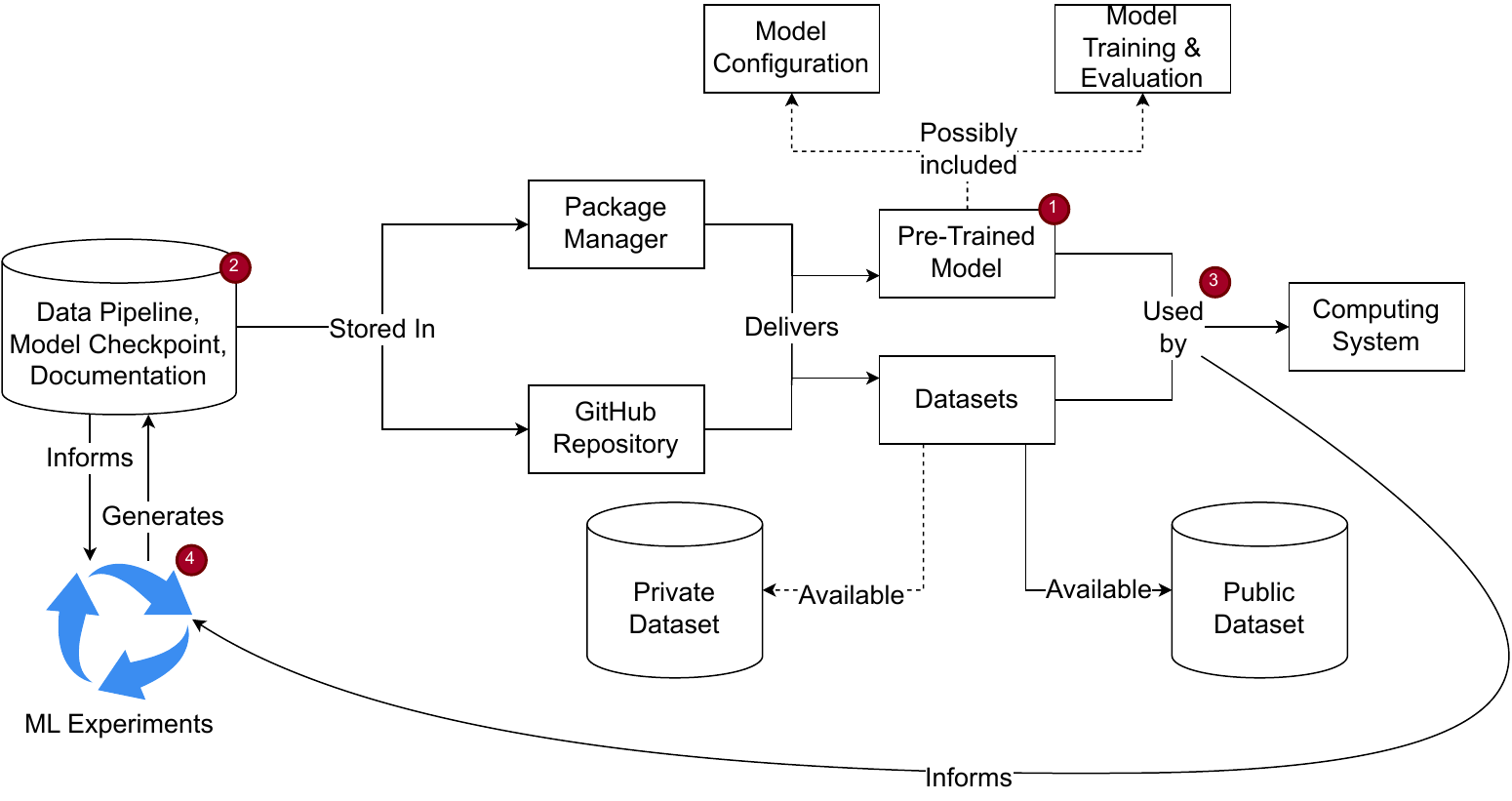}
    \caption
    {
    This figure illustrates the complexity of mining and using DNN packages.
    We note four challenges in mining DNN packages.
    (1) A DNN package may solely contain the pre-trained model with no additional information, limiting what can be learned from it.
    (2) ML experiments generate new data pipelines, model checkpoints, and documentation as they are iterated upon and improved on. It is challenging to describe and learn from the interrelationships of the experiments. 
    (3) ML experiments are affected by pre-trained models trained on similar tasks or datasets, or by the dataset itself. There is a cyclical dependency that affects mining these artifacts.
    (4) ML experiments may not be made publicly available. This ``secret''~\cite{Aranda2009SecretLifeofBugs} process is crucial to deep learning engineering but difficult to observe.
    }
    \label{fig:miningChallenges}
\end{figure*}


As with any other software engineering effort, conceptualizing, adapting, and deploying DNNs is a measurable process.
A high-quality software artifact is typically associated with an effective software engineering process~\cite{sommerville2015software}.
By measuring the software engineering process of DNNs, engineers can gain insights into the quality of the DNN.
These process measurements can assist engineering teams in evaluating DNNs as dependencies within their projects.

There are no tools that quantify what is and is not an effective software engineering process tailored to DNNs.
Existing understanding~\cite{fenton_software_2014} and tooling~\cite{synovic_snapshot_2022} can be applied to the DNN engineering process.
However, as the reuse, conceptualizing, adapting, and deploying process of DNNs is different than traditional software projects, it is unclear as to what is and is not a good engineering process for DNNs.

\cref{fig:miningChallenges} illustrates the challenges of measuring the DNN engineering (and reuse) process, suggesting the difficulties in mining to elicit empirical data about effective processes.
Unlike traditional software engineering packages, DNN packages cannot be analyzed in isolation;\footnote{Of course, traditional software should also not be analyzed in isolation, but more useful data may be derived from an isolated view in that case.} the training dataset, config, and documentation all provide crucial information as to both what the model is, as well as the performance is.
Due to the additional information, the data extraction that is required to mine a DNN repository is more complex. 

To assist in the mining efforts of DNNs, the PTMTorrent dataset~\cite{jiangPTMTorrentDatasetMining2023} provides a dataset of $\sim$60,000 PTMs from 5 model registries.
This dataset contains the full \code{git} repository for each DNN package, including the model, documentation, and configuration information.
The follow-up PeaTMOSS dataset adds connections to use in open-source projects from GitHub~\cite{jiang2024peatmoss}.

\section{Discussion} \label{sec:discussion}




\subsection{DNN Reuse as an Accelerator}

Reuse is a key engineering technique because it offers such substantial cost savings~\cite{jones2011economics}.
Whether a software engineer reuses via the conceptual paradigm, the adaptation paradigm, or the deployment paradigm (or all three!), they benefit.



Our purpose in framing deep learning engineering in terms of reuse is to emphasize the potential benefits of deep learning reuse to software engineers.
However, we acknowledge that at present the benefit may not always be realized.
For example, in recent interviews we found that skilled software engineers sometimes take months to identify a good model to adapt and fine-tune to reach acceptable performance (adaptation paradigm)~\cite{Jiang2022PTMReuse}.
We are not aware of any direct comparisons in engineering cost between from-scratch development and reuse development.
However, we expect that advances in DNN reuse techniques can substantially lower the current costs.
We challenge the research community to show the way forward.

\subsection{DNN Reuse in the Age of Foundational Models}


The trend in deep learning has been to develop larger and larger models, \eg supporting multiple modes of input and capable of solving a wide range of problems.
The resulting models are generalists.
One might wonder, when a single model like GPT-4 can solve math problems and interpret jokes~\cite{bubeck2023sparks}, why researchers and engineers should be concerned with model reuse as opposed to simply model application.
While large-language models at the time of writing are appearing to become less open as part of an industry ``arms race'' of sorts, investigating reuse is important regardless of whether building proprietary and/or open source deep learning solutions (similar to the corresponding situation in general software development).
We point out two weaknesses of the current trend towards a single powerful model.

First, this approach centralizes power in the hands of those who control the model~\cite{zhang2022opt}.
Foundational models such as ChatGPT and GPT-4 are trained by one organization with unclear data sources, and they impose a client-server model where one organization mediates all responses.
Centralization and unmonitored data processing bode poorly for user privacy~\cite{sun2022coprotector} and the concomitant problems of a monopoly~\cite{spence1975monopoly}.

Second, this approach is highly energy-consumptive~\cite{strubell2019energy}.
Resource-intensive models can achieve state-of-the-art performance, but sometimes this is not necessary~\cite{you2023zeus}.
In resource-constrained environments, such as those involved in real-time decision-making, fast and adequate may suffice.
In this context, the reuse paradigms allow engineers to explore alternative approaches that trade off resource utilization and performance.




\section{Conclusion} \label{sec:conclusion}

This paper described the challenges of re-using deep neural networks (DNNs). 
DNNs have demonstrated exceptional performance in various domains, but their development and computational costs remain significant.
DNN re-use can reduce development costs.
However, re-use presents its own set of challenges, encompassing technical capabilities and engineering practices.
We hope this work improves understanding DNN re-use.
We discussed three types of reuse --- conceptual, adaptation, and deployment.
By identifying the gaps in current re-use practices, we contribute to the understanding of DNN reuse and offer insights for future research in this area.





\section*{Acknowledgments}

We acknowledge financial support from
 Google,
 Cisco,
 and
 NSF awards 2229703, 2107230, 2107020, and 2104319. 

\balance

\bibliographystyle{IEEEtran}
\bibliography{bib/final,bib/extra-bib}

\end{document}